\documentclass[twocolumn,showpacs,preprintnumbers,amsmath,amssymb,prb]{revtex4}

\usepackage{graphicx}% Include figure files
\usepackage{subfigure}
\usepackage{bm}% bold math

\begin{document}

\preprint{}

\title{Temperature dependence of the charge carrier mobility in gated
quasi-one-dimensional systems}
\author{Lazaros K. Gallos}
\affiliation{Department of Physics, University of Thessaloniki, 54124, 
Thessaloniki, Greece}

\author{Bijan Movaghar}
\affiliation{SOMS Centre, Department of Chemistry, University of Leeds, Leeds 
LS2-9JT, United Kingdom}

\author{Laurens D.A. Siebbeles}
\affiliation{Interfaculty Reactor Institute, Delft University of Technology,  
Mekelweg   15, 2629 JB Delft,  The Netherlands  }

\date{\today}% It is always \today, today,

\begin{abstract}
The many-body Monte Carlo method is used to evaluate the frequency dependent 
conductivity and the average mobility
of a system of hopping charges, electronic or ionic on a one-dimensional chain 
or channel of finite length. Two cases
are considered: the chain is connected to electrodes and in the other case the 
chain is confined giving zero dc conduction.
The concentration of charge is varied using a gate electrode. At low 
temperatures and with the presence of an injection barrier,
the mobility is an oscillatory function of density. This is due to the 
phenomenon of charge density pinning. Mobility changes occur
due to the co-operative pinning and unpinning of the distribution. At high 
temperatures, we find that the electron-electron interaction
reduces the mobility monotonically with density, but perhaps not as much as one 
might intuitively expect because the path summation favour
the ``in-phase contributions'' to the mobility, i.e.  the sequential paths in 
which the carriers have to  wait for  the one in front to exit
and so on.  The carrier interactions produce a frequency dependent mobility 
which is of the same order as the change in the dc mobility with
density, i.e. it is a comparably weak effect. However, when combined with an 
injection barrier or intrinsic disorder, the interactions 
reduce the free volume and amplify  disorder by making it non-local  and this  
can explain the too early onset of frequency dependence
in the conductivity of some "high mobility quasi-one-dimensional organic 
materials.
\end{abstract}


\maketitle

\section{Introduction}

The purpose of this paper is to understand how electron-electron interactions 
affect the ac conductivity in the stochastic regime of transport in low 
dimensional materials. Transport in one-dimensional or quasi one-dimensional 
molecular wires formed with self-assembled discotic 
molecules\cite{1a,1b,1c,1d,2a,2b,3}, nanotubes\cite{4,5} at high temperatures or with disorder,
and MBE gate engineered confined charged channels, constitutes currently a very hot and 
technically important topic\cite{6}. Other important systems include 2-
dimensional layers such as those encountered in smectic liquid 
crystals\cite{7,8}, gate charge injected or doped polymers\cite{9}, gate 
addressed FET devices based on organic materials \cite{10}, thin film 
transistors, and same such systems when they exhibit a metal insulator 
transition and superconductors. 

One-carrier conduction phenomena are now reasonably well understood, and many 
workers are in the process of designing structures in order to exploit the 
subtle phenomena associated with one-carrier resonant tunnelling through large 
molecules \cite{12a,12b,12c}. The molecules have energy levels structure which 
can be put in and out of resonance to the energy of the incoming waves. This can 
be done using gates and nano-gates. The changes in these energy levels caused by 
contamination or interaction or external fields can then be recorded as a new 
current-voltage response and used in various device applications. At the same 
time it is also well known that small low-dimensional structures are more 
susceptible to electron-electron interactions. As long as one is dealing with a 
few steady state carriers which can escape before others arrive, mean-field 
approaches are in order. Simple minded mean-field approaches will however for 
obvious reasons not work in low-dimensional systems in which carriers interfere 
seriously with each others pathways. Interactions are known, for example, to 
play a role in carbon nanotubes at low temperatures, and we believe that many ac 
response phenomena in conjugated polymers and molecular wires have in the past 
been wrongly attributed to pure disorder effects. In reality one is often 
dealing with a combination of both electron-electron interactions and disorder 
and it is not easy to differentiate the two experimentally unless one has the 
right theoretical tools. Electronic interactions also come into play with high 
charging, such as occur for example in electrode or gate charged FET organic 
interfaces\cite{10} and thin film transistors. The same applies to highly doped 
materials. Charging energy is particularly important in high-band gap, low 
dielectric constant and low-dimensional molecular materials such as chemically 
or injection doped smectic and discotic liquid crystals\cite{1a,1b,1c,1d,7}.

Charges can be injected into organic materials chemically or by high fields and 
by light and one can measure the frequency dependent conductivity. The frequency 
dependent conductivity is in principle one of the easiest ways of seeing the 
effect of interactions.  The long range coulomb forces produce, naively speaking, 
disorder-like potentials which confine and scatter the carriers. They 
generate in principle a frequency dependence on the motion which looks similar 
to the one produced by true disorder. The problem in practice is, however, that 
one always has some disorder and some electrode polarisation effects, and then 
it is difficult to disentangle these processes from each other.

\begin{figure}
\includegraphics{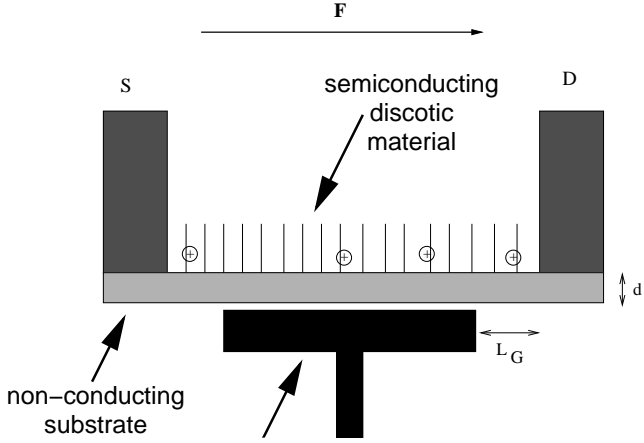}
\caption{\label{fig1}Schematic diagram of the gated model system. A molecular 
wire is first
charged by a gate, and the conduction takes place along the molecular column 
between the source (S)
and the drain (D).}
\end{figure}

\section{The computational model}

Using many particle Monte-Carlo simulations, we study the effect of 
charge-charge interactions on the ac conductivity of finite and electrode 
addressed one-dimensional chains following the work in reference 
\onlinecite{13}. A schematic representation of the system is shown in figure 
\ref{fig1}. We consider that a molecular columnar wire, which does not interact 
with its neighboring columns, is charged via a gated electrode and is attached 
to two metal electrodes, which serve the purpose of source and drain. The gate 
electrode lies at a distance of $d$=1 nm from the column, and its distance from 
the edge electrodes equals $L_G$=1.4 nm. We have also considered the more 
realistic situation that the gate electrode is $d$=30 nm away from the column. 
A number $N$ of charges (negative or positive, in our case negative) is placed
along this linear gate, which is parallel to the organic column. 
Charges (which can be either holes or electrons, and in our case we consider them
to be holes) are created in the column and move along its length under the influence of 
an externally applied field, which can be either DC or AC. These holes can enter or 
leave the column through the metal electrodes, but asymptotically they establish a steady-state 
distribution in the column.

In practice, we consider a column of 300 sites, with a lattice constant $a=0.35$ nm,
which corresponds to a $0.1$ micron device. The number $N$ of charges in the gate is
fixed by the assumed value of the gate field near the plate right at the start
and remains constant throughout the simulation. We assume that the negative voltage at the gate causes an equal 
number of charges (holes) to appear in random positions in the column.

The Coulomb energy $V_n(t)$ of a hole located on site with index $n$ at distance $r_n=na$
from the source electrode,
is the sum of repulsive forces due to the presence of other holes and the attractive
force due to the negatively charged gate:
\begin{equation}
V_n(t) = V_{n,{\rm rep}}(t) + V_{n,{\rm att}} = \frac{e^2}{4\pi \epsilon_0 \epsilon_r} \sum_{m=1}^{L} \frac{H_m(t)}{|r_n-r_m|}
+ V_{n,{\rm att}}\;,
\end{equation}
where $H_m(t)$ is the hole density of site m at time $t$, and assumes the values of 1 or 0, depending
on whether this site is occupied by a hole at time $t$.

For the calculation of the Coulomb attraction energy on the holes we use the jellium model.
We consider that the number of charges in the linear gate are $N$, the length of the column is
$L$, the gate electrode starts at a distance $L_G$ from the source and drain electrodes and 
its distance from the column is $d$. Under these conditions, the attraction energy for a hole
located on site $r_n$ is equal to
\begin{widetext}
\begin{equation}
V_{n,{\rm att}} = \left\{
\begin{array}{l}
\frac{e^2}{4\pi\epsilon_0 e_r} \frac{N}{L-2L_G}
\ln\left(\frac{L-L_G-r_n+\sqrt{(L-L_G-r_n)^2+d^2}}{L_G-r_n+\sqrt{(L_G-r_n)^2+d^2}}\right), r_n<L_G \\

\frac{e^2}{4\pi\epsilon_0 e_r} \frac{N}{L-2L_G}
\ln\left(\frac{\left( r_n-L_G+\sqrt{(L_G-r_n)^2+d^2}\right) \left( L-L_G-r_n+\sqrt{(L-L_G-r_n)^2+d^2}\right)}{d^2}\right), L_G\leq r_n\leq L-L_G\\

\frac{e^2}{4\pi\epsilon_0 e_r} \frac{N}{L-2L_G}
\ln\left(\frac{r_n-L_G+\sqrt{(r_n-L_G)^2+d^2}}{r_n-L+L_G+\sqrt{(r_n-L+L_G)^2+d^2}}\right), r_n>L-L_G\\

\end{array}
\right.\;.
\end{equation}
\end{widetext}

During each step, 
the transition probabilities for all the charges in the column are computed, so 
that a charge in site $n$ has a hopping transition rate $W_{n,n\pm 1}$ towards 
its neighbors
\begin{equation}
W_{n,n\pm 1} = \left\{ 
\begin{array}{cc}
\nu_0 \exp\left( -\frac{\Delta E_{n,n\pm 1}}{kT} \right) &  {\rm for} \Delta 
E_{n,n\pm 1}>0\\
\nu_0 & {\rm for} \Delta E_{n,n\pm 1} \leq 0
\end{array}
\right.\;,
\end{equation}
The energy difference between the two sites is computed by the formula
\begin{equation}
\Delta E_{n,n\pm 1}=V_{n,n\pm 1}-V_n\mp eFa \;,
\end{equation}
where the symbol $V_{n,n\pm 1}$ represents the Coulombic energy on site $n\pm 1$ when we displace the hole located on
site $n$ to the site $n\pm 1$.
The jump frequency $\nu_0$ 
is fixed to 10$^{12}$ Hz. The external electric field $F$ can either be constant $F_0=$20 
kV/cm, or oscillating with a period $\omega$ so that $F(t)=F_0 \cos(\omega t)$.
When a neighboring site is occupied the transition rate towards that site is 0, i.e. we
use excluded volume rules.

For the first and last site of the column the quantity $\Delta E$ is modified, 
since now there is only one neighbour to move to, and injection/absorption to 
the electrodes is possible. There exists an energy barrier between the valence band of the organic material
and the Fermi level of the metal electrode, whose height we denote by $E_b$.
The energy difference for a hole injection from the electrode to the first column site is
\begin{equation}
\Delta E^{\rm inj}_{n=1}=E_b + V_1-eFa \;,
\end{equation}
and for hole injection from the other electrode to the opposite edge is
\begin{equation}
\Delta E^{\rm inj}_{n=L}=E_b + V_N+eFa \;.
\end{equation}
In the same way, the energy difference for absorption of a hole at $n=1$ to the adjacent electrode is
\begin{equation}
\Delta E^{\rm abs}_{n=1}= -E_b - V_1 + eFa \;,
\end{equation}
while the same quantity for a hole leaving the column at the other side is given by
\begin{equation}
\Delta E^{\rm abs}_{n=L}= -E_b - V_N - eFa \;.
\end{equation}
The case of $E_b=\infty$ corresponds to blocking barriers, where the number of holes in the
system is fixed for the entire duration of the simulation and no hole can enter or leave the column.

Once the transition probabilities $W$ have been computed for all holes in the 
system, the time $\delta t$ required for each hop is drawn by an exponential 
distribution,
\begin{equation}
\delta t = \frac{-\ln R}{W_{n,n-1}+W_{n,n+1}} \;,
\end{equation}
with $R$ being a uniformly distributed random variable, assuming values between 0 and 1. The shortest time $\delta t_{\rm min}$ of 
this ensemble (including the time needed for injection/absorption) is chosen and 
this jump takes place. The total time advances by the same amount $\delta t_{\rm min}$, and the 
direction of the jump is decided with a probability
\begin{equation}
P_{n,n \pm 1} = \frac{W_{n,n \pm 1}}{W_{n,n-1}+W_{n,n+1}} \; .
\end{equation}
When a hole moves from a site to another, the work done by the electric field during the time
interval $\delta t_{\rm min}$ is calculated by $\delta A = \pm eaF(t)$. The total work done over
a time interval is added to the total work $\Delta A$ for that interval.

The above procedure is repeated many times with the holes moving around, leaving 
and entering the column. The time that a given realization of the system lasts 
depends on the convergence of the results, which in turn depends on the specific parameters of
the problem. Thus, typical times simulated are in the range of 0.01-0.1 ms.
In order to eliminate any initial distribution bias, we 
repeat the whole algorithm for a number of different realizations (typically of the
order of 10-100), with different random initial configurations. During all these realizations we 
monitor the average of the holes mobility $\mu$ in the system with time, until 
this average is well converged. In almost all cases, the error bars for the mobility
estimation were less than 10\%.

The number of holes $N_h$ in the column is a number which varies with time and
converges asymptotically to an average value, depending on the exact simulation conditions.
When we use blocking electrodes, of course, $N_h$ is always equal to $N$, the
number of negative charges in the gate.

The average hole mobility $\mu$ is calculated via the work $\Delta A$ done on 
all holes during an oscillation period $\Delta t =2\pi / \omega$, and is given 
by\cite{13}
\begin{equation}
\mu = \frac{2\Delta A}{e F_0^2 \Delta t N_h} \;.
\end{equation}

\section{Results}

We have simulated the gated system described above for both DC and AC externally 
applied field. When we considered an AC field, we also studied the following two 
cases: a) include a finite barrier height between the metal electrodes and the 
column, and b) assume an infinite barrier height, which corresponds to blocking 
electrodes.

\subsection{DC Mobility}

\begin{figure}
\includegraphics{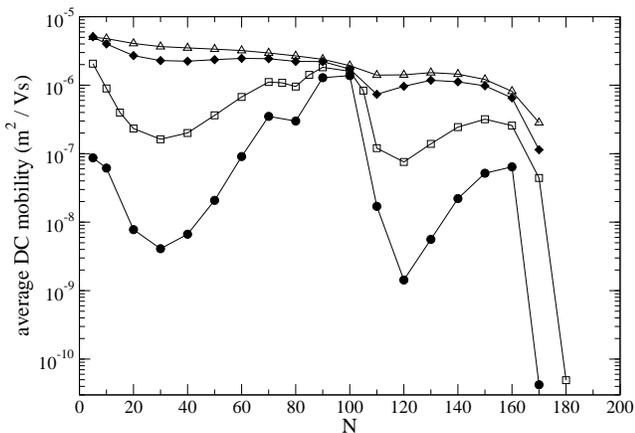}
\caption{\label{fig2}Average carrier mobility as a function of the number of  charges 
$N$ in the gate for a DC field of 20 kV/cm.
Different temperatures are plotted(top to bottom): 450 K, 300 K, 150 K, and 100 
K. Lines are used as guides to the eye.}
\end{figure}

In figure \ref{fig2} we present the average mobility for the gated 
one-dimensional column with 300 sites in a 
DC field $F$=20 kV/cm, as a function of the number of charges $N$ in the gate. 
The injection energy barrier height is equal to $E_b$=0.21 eV.
Neglecting all carrier-carrier interactions, including those from the gate charge,
the mobility would scale as
\begin{equation}
\mu = \mu_0 \exp\left( -\frac{E_b}{kT} \right) \;.
\label{EQmob1car}
\end{equation}
Different curves 
correspond to runs at different temperatures. Figure 2 shows that the temperature dependence of the 
mobility is strongly dependent on the value of $N$, the charge density. The 
increase in mobility with temperature is due to two factors. a) The carriers find 
it easier to come into the chain, and as the temperature increases because of the
barrier factor (i.e. as in equation \ref{EQmob1car}) b) the charges 
can also hop more easily inside the chains against the self-field produced by
the other carriers, which means that in equation \ref{EQmob1car} the factor
$\mu_0$ should be replaced by $\mu_0(T,N)$. The calculated low 
temperature mobility is very small at $T$=50 K (not shown). In this limit
the charges are 
practically immobile and cannot overcome the charge-charge interactions which 
keep them vibrating around their ``equilibrium'' positions. 

An extremely interesting phenomenon is observed when we look at the temperature 
dependence of the calculated mobility at different densities. Figure \ref{fig2} shows
that there are densities at which the ``activation energy'' is practically zero. 
This means that for these particular values of $N$, a charge distribution is 
established in the given field geometry, where the injection barrier is 
compensated by the internal space charge and the gate field. One can also look at it
as the variation of the calculated mobility with the number of charges $N$. When the temperature is 
constant, the average mobility decreases with increasing $N$. This is expected 
because  the carriers have less space available to move in. This observation is 
actually true for room temperatures and above, but figure \ref{fig2} shows that
this rule is clearly violated at lower 
temperatures, and for high concentrations. For low temperatures, the mobility 
decreases with charge concentration, but when the concentration exceeds a 
certain value (around $N$=30, corresponding to a concentration of 0.1) the 
mobility increases with increasing concentration until $N$=100. The mobility 
decreases again, but exhibits another peak at $N$=150. For concentrations higher 
than 0.5 the average mobility drops extremely rapidly, and the charges are 
practically immobile. In other words, in the mobility versus $N$ plot, we can observe three maxima 
(located roughly at $N$=5, $N$=100, and $N$=150) and three minima (for $N$=30, 
$N$=120, and $N$=180, which is the highest concentration we used). One can 
understand this behavior if one notes that the combination of the intrinsic (work function)
barrier, the charging energy coming from other carriers, and the charging energy from
the gate within a particular gate geometry, together produce a density dependent 
pinning barrier which modulates the absorption and injection efficiency.
One can see that when absorption/injection efficiency is high, the mobility can be
high, and this is caused by the forward drift of all carriers. As soon as the density
exceeds the critical number $N_c$ the injection is reduced  and the incoming carrier
has to wait until one
has exited . To calculate $N_c$  we note that the  Coulomb barrier at
entry on column is defined by
\begin{equation}
E_{\rm CB} = \sum_{n=1}^N \frac{e^2}{4\pi \epsilon \epsilon_0 n \langle a(N) \rangle} - E_{\rm gate} \;,
\end{equation}
where $\langle a(N) \rangle$ is the average distance between two carriers at total number $N$. With
$\epsilon \sim 3$, $E_{\rm CB}$ exceeds $kT$ when the number of carriers $N$ on the
chain is larger than $N_c\sim 40-50$ (the value increases slightly with the temperature).

\begin{figure}
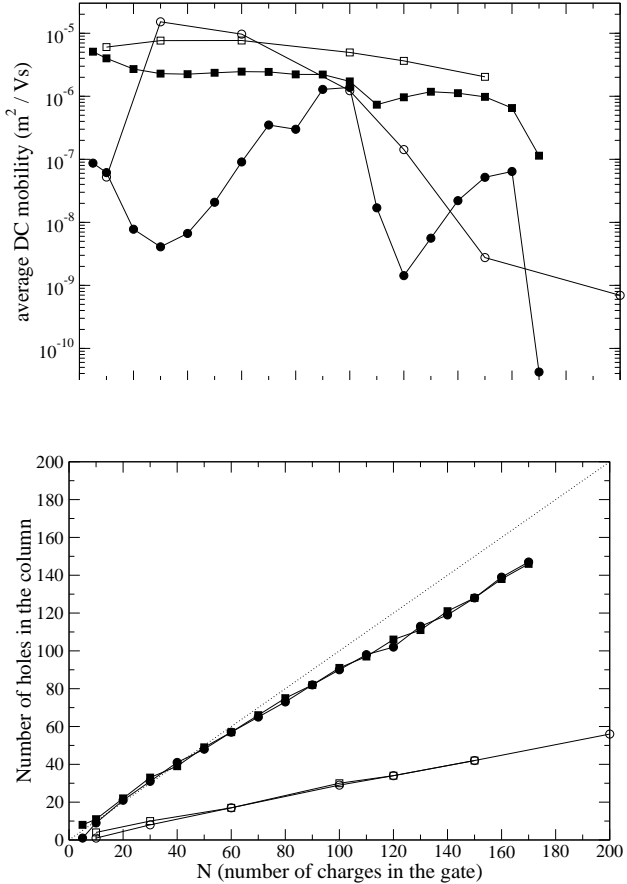

\subfigure{\includegraphics{fig3a.eps}}
\includegraphics{fig3b.eps}
\caption{\label{fig3}Comparison between systems with $d$=1nm (filled symbols) 
and $d$=30 nm (open symbols) of a) the average mobility and
b) the average number of charges in the column, for $T$=100 K (circles) and 
$T$=300 K (squares), under the influence of a DC field of 20 kV /cm.}
\end{figure}

When the Coulomb repulsion energy at the entry site starts to get much bigger than
$kT$ the carriers block each other along the column. They can only move sequentially,
``in a phase'' forced on them by the internal fields. The total energy barrier seen by the carrier
on the first site  in the column is given by
\begin{equation}
E_{\rm tot} = E_b + \sum_j \frac{e^2}{4\pi \epsilon \epsilon_0 r_{ij}} - E_{\rm gate} \;,
\label{Etotal}
\end{equation}
where $E_{\rm gate}$ is the ``attractive'' energy of the gate counterions and depends on
the distance and gate geometry, and the sum runs over the position of the carriers.
When we increase the distance to the gate counter charge distribution from 1 nm
to a more 
realistic value for conventional gate devices of $d$ = 30 nm (see figure 
\ref{fig3}) \cite{10,14a,14b}, the mobility versus density behaves quite differently
as shown in figure \ref{fig3}. 
The internal charge density as a function of gate voltage is also shown in 
figure \ref{fig3}. Note that we now have fewer charges for the same gate charge 
because the coulomb repulsion is relatively stronger and prevents charges from
coming into the column. Note also that the mobility versus $N$
oscillations observed in figure \ref{fig2} no longer appear in the same place. 

For higher temperatures, the picture is much easier to understand because the 
charges have enough energy to overcome the local attractive energy barriers and there is no 
pinning effect. When $T$=450 K and $N$=5, for example, it is much easier for 
holes to enter the system, so that there are always 8 to 9 charges in the column 
which undergo a forward drift motion, as opposed to 1 or 2 charges in the column 
for $T$=100 K.

For the injection barrier $E_b$ value used, the average number of holes in the 
column during the simulation is close to the value expected from the gate 
charge, except of course when we use $d$=30 nm. Here, the Coulomb repulsion with 
weak screening keeps the charges out of the column (see figure \ref{fig3}). For 
$d$=1 nm, the gate to charge density matching is almost one to one for all 
temperatures, and at the lower end of concentrations. At higher concentrations, 
we observe a reduction in the average number of holes in the column of the order 
of 10-15\%, e.g. for $N$=100 we have around 90 holes. This is not surprising and 
is an effect of the ``not metallically screened coulomb repulsion'' in the chain 
\cite{14a,14b}. 

\begin{figure}
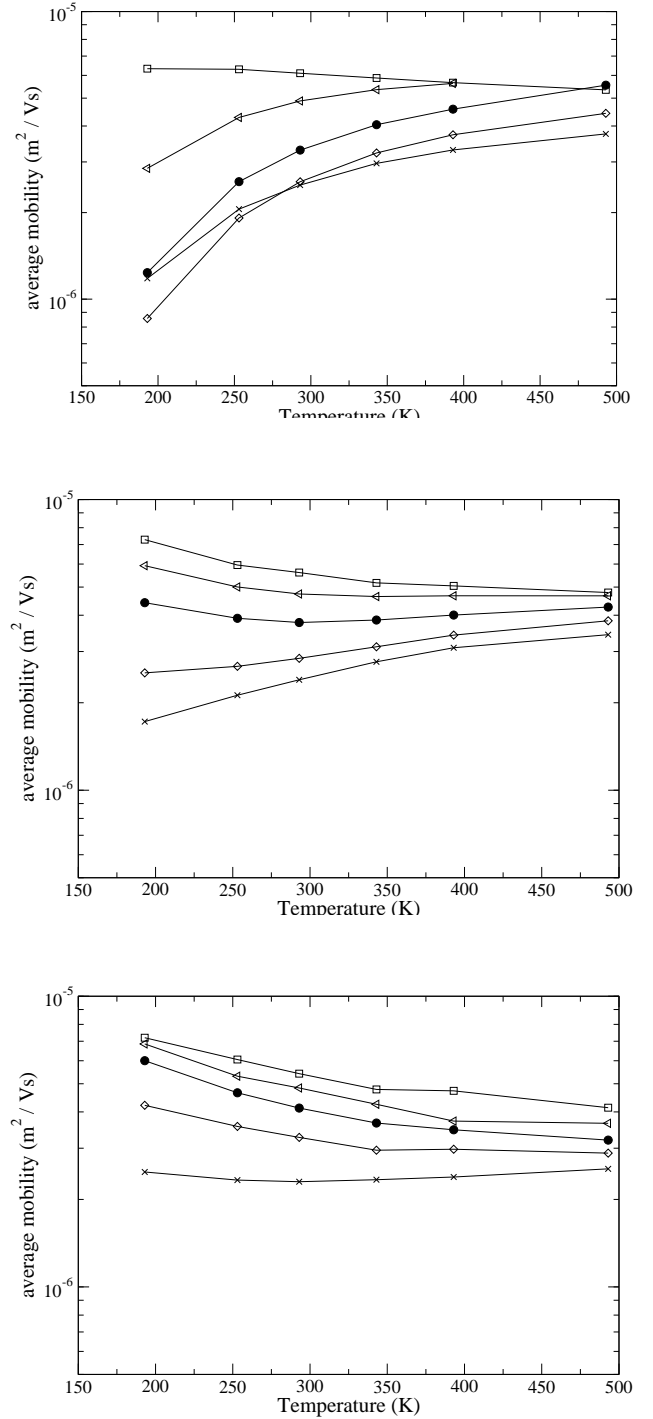

\subfigure{\includegraphics{fig4a.eps}}
\vspace{.4 cm}
\subfigure{\includegraphics{fig4b.eps}}
\includegraphics{fig4c.eps}
\caption{\label{fig4}Temperature dependence of the average AC mobility for 
different initial charge concentrations and field frequencies: a) 10$^6$ Hz, 
b) 10$^9$ Hz, and c) 10$^{10}$ Hz. The energy barrier height of the electrode is 
equal to 0.21 eV. We observe different patterns of behavior as a function of the 
concentration as we vary the frequency. The gate charges are as follows: $N$=3 
($\Box$), $N$=9 ($\triangleleft$), $N$=18 ($\bullet$), $N$=36 ($\Diamond$), and 
$N$=60 ($\times$).}
\end{figure}

\subsection{AC Mobility}

\subsubsection{Finite electrode barrier height}

For the relatively low frequencies of 10$^6$ Hz (figure \ref{fig4}a), and for 
intermediate densities, the mobility increases almost one order of magnitude as 
we increase the temperature (the injection barrier is limiting the mobility). 
But, at the low concentration end,
when carrier interactions are small the temperature dependence is weak.
At very low frequencies, one should of course expect similar
results to the ones observed in the limit of a DC field. So, the temperature independence
at low concentrations
implies that at 10$^6$ Hz one is already looking at
bulk motion. The barrier no longer affects the ``single-particle'' mobility.
It is important to note that apparently all the temperature dependence
in figure \ref{fig4}a is already due to many-body
effects. When the 
frequency of the electric field is increased further, as shown in figure \ref{fig4}b,
to the frequency value of 10$^9$ Hz for example, the change with temperature is now
weaker even for higher densities. On this time scale, the interactions are getting less
important as one might expect. Indeed, 
for very high frequencies, e.g. 10$^{10}$ Hz as shown in 
figure \ref{fig4}c, the mobility actually decreases with increasing temperature, 
and this behavior is monotonic with $N$. The change in mobility is however less than
half an order of magnitude. This behavior is unusual for hopping systems\cite{15}.
We believe that this is because  the effective disorder is increasing with temperature.
The disorder increases because the number of carriers in the column is increasing and this reduces
the free volume, as shown in figure \ref{fig2}.

\begin{figure}
\includegraphics{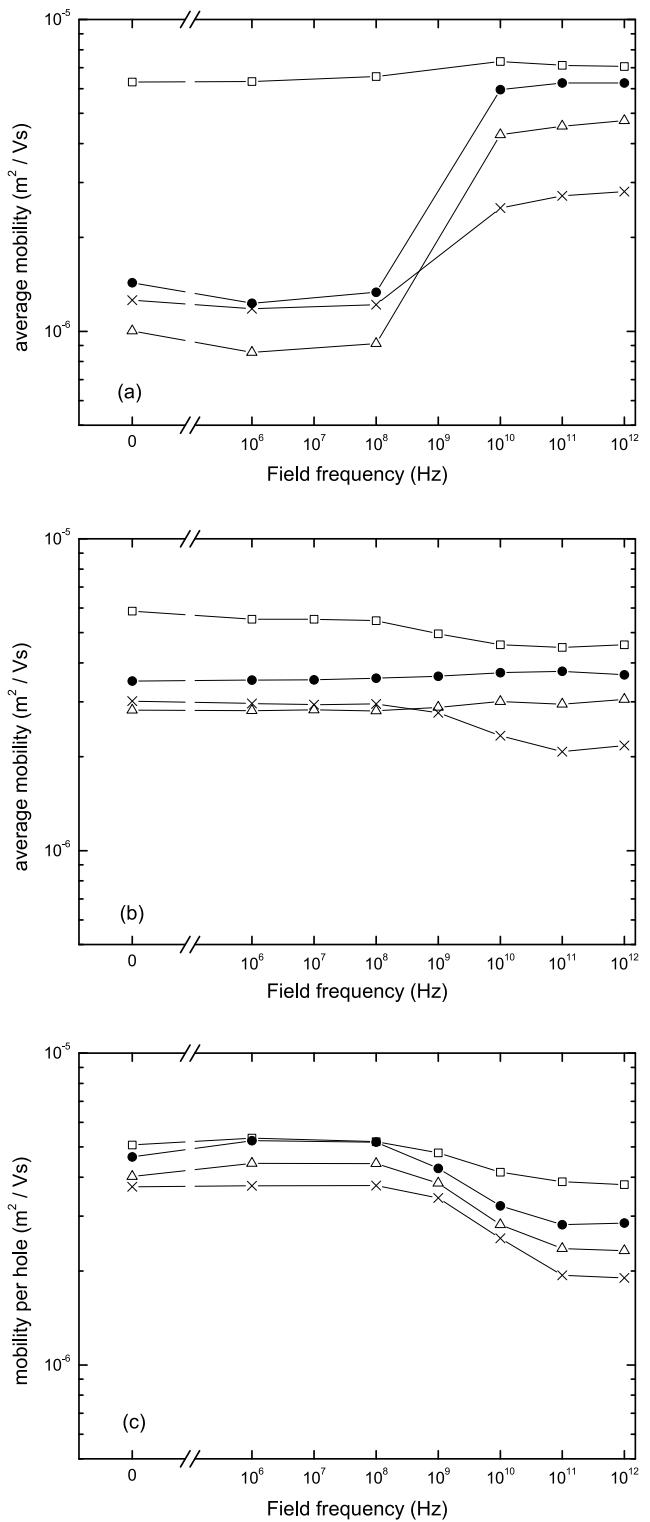}
\caption{\label{fig5}Variation of the average AC mobility with the field 
frequency. The energy barrier height of the electrode is equal to 0.21 eV. Three 
different temperatures are presented: a) T=193 K, b) T=343 K, and c) T=493 K. 
Symbols correspond to: $N$=3 ($\Box$), $N$=18 ($\bullet$), $N$=36 
($\bigtriangleup$), $N$=60 ($\times$).}
\end{figure}

\begin{figure}
\includegraphics{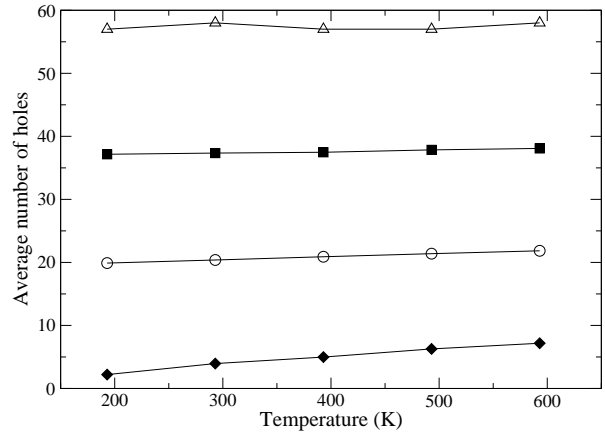}
\caption{\label{fig6}Average over all AC frequencies of the number of holes 
along the column, as a function of temperature. The energy barrier height of the 
electrode is equal to 0.21 eV. Top to bottom: $N$=60, $N$=36, $N=18$, and $N=3$. 
Note the charging effect at higher temperatures and lower concentrations.}
\end{figure}

For a given temperature, the calculated frequency dependence is shown in figure \ref{fig5}. 
Here we have plotted 3 curves: a) $T$=193 K, b) $T$=343 K, and c) $T$=493 K. Consider 
first the low temperature curve. Here the average mobility at small concentrations is 
practically constant for all frequencies. This means that the injection barrier 
has very little effect on the frequency behavior at this temperature.
Let us try to understand why.
In the spirit of a simple resistance sum\cite{16} for a one particle random walker and
rigorous in the low frequency limit, we have for a column of 300 sites a transport
admittance $Y$:
\begin{equation}
Y(\omega) = \frac{e^2 a^2}{kT} \left( \frac{1}{i\omega + W_{\rm inj}} +
\frac{299}{i\omega + W_{\rm bulk}} \right)^{-1} \;.
\label{resist}
\end{equation}
The injection rate $W_{\rm inj}$ is activated as in equation \ref{EQmob1car}, and
the bulk rate $W_{\rm bulk}$ is a constant for a single particle.
From equation \ref{resist} one would  expect the
frequency dependence to come in when the frequency is of the order of
the injection rate, that is roughly 10$^8$ Hz, and to saturate at the
bulk value of 10$^{12}$. This, however, does not happen in figure \ref{fig5}.
At $N=3$ the mobility stays roughly at saturation value all the way. The
reason for this behavior is as shown in figure \ref{fig2}. The particles  in the
simulation are moving in a potential made up of the gate field, the
electron-electron interaction and the barrier, i.e. as given by equation \ref{Etotal}.
The barrier at a given  concentration may indeed be much lower  than
the injection  barrier,  and that is why there is no, or only a weak
frequency dependence for this particular concentration. Looking at
figure \ref{fig5} we see that  at higher concentrations the mobility becomes
strongly frequency dependent (see the corresponding  dc limit  in
figure \ref{fig2}), and rapidly increases  with frequency at around 10$^9$ Hz.  This is
now an interaction effect as in equation \ref{Etotal},  and it looks as if  in the spirit
of one body theory,  interactions are  causing an effective disorder.
The interactions  appear to have   little effect  on  time scales
shorter than $10^{-9}$ s. As we raise the temperature further to 343 K,
the frequency dependence is now  weak for all $N$.  Above  343 K, there is
in effect no strong inhomogeneity in the  hop time  distribution. This
means that the  variation in hopping activation energies is comparable
to $kT$ . Finally,  in the extreme high $T$ limit, when $T$=493 K the mobility
actually  decreases with increasing frequency. This is most  unusual for
hopping transport and is  in this model due to the extra charging that
high temperature causes, as shown in figure \ref{fig2}.
At high densities, a real time analysis shows that  
the charges move like a  collective ``line of billiard balls''. A carrier can 
only come in when one has escaped, and the mobility decreases with $N$ at high  
temperatures. In figure \ref{fig6}, we present the steady 
state number of holes in the column as a function of temperature for different 
gate charges. This number is an average over all different frequencies used. It 
is evident that there is a considerable net charging effect in the system. The 
charging is more prominent for smaller $N$. For example, when $N$=3 and $T$=593 
K the average number of holes in the system is around 7, while for $T$=193 K it 
is 2 (the absolute value of these numbers depends of course on the value of the 
interface energy barrier $E_b$, but the basic trends remain the same). This 
charging effect is responsible for giving the lower value of the mobility 
observed at higher temperatures (see for example figure \ref{fig2}). Normally, 
in hopping transport we expect the mobility to go up with 
temperature. Here however, the steady state density of holes, which also
influences the magnitude of the hole-hole scattering, is 
temperature-dependent. There are competing influences: whereas temperature
makes hopping easier, the number of holes also goes up, and this
increases therefore the hole-hole interactions, or what is also the 
degree of disorder in the system.

\subsubsection{Blocking electrodes}

\begin{figure}
\includegraphics{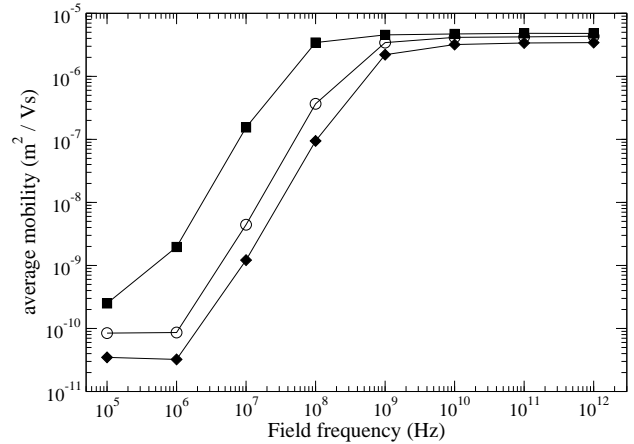}
\caption{\label{fig7}Average mobility as a function of frequency for different 
charge concentrations (top to bottom: $N$=3, $N$=18, $N$=36 ) with blocking 
electrodes and 300 sites at T=293 K. }
\end{figure}

We also studied the case where there is an infinitely high energy barrier among 
the electrodes and the molecular wire. This corresponds to blocking electrodes, 
which means that the system is ``closed'' in the sense that no carrier can 
(subsequent to charging) enter or leave the column. In figure \ref{fig7} we can 
see that at high frequencies the average mobility is constant and almost the 
same for different concentrations. This is because in this limit we are looking at
bulk hopping. As we lower the frequency, the mobility drops 
steadily to very small values. The carrier displacement is 
at low frequencies limited by the boundary and the decrease in mobility
is expected in a confined system. The effect 
is also concentration dependent. Going up in frequency, we
observe that the higher the hole densities, the later the saturation
frequency. For the same frequency, the difference in the mobility for
the low and high concentrations can be up to 2 orders of magnitude. This is true for 
concentrations in the range from 0.01 to 0.12 charges per site.

\begin{figure}
\includegraphics{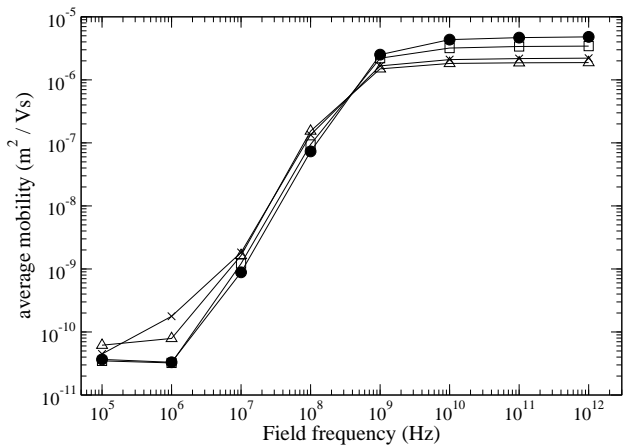}
\caption{\label{fig8}Average mobility as a function of frequency for different 
temperatures with blocking electrodes and 300 sites, where N=36.
Symbols correspond to the following temperatures: $T$=193 K ($\bullet$), $T$=293 
K ($\Box$), $T$=493 K ($\times$), and $T$=593 K ($\bigtriangleup$).}
\end{figure}

The corresponding temperature/frequency dependence is shown in figure \ref{fig8}. The curves 
seem to lie practically all on the same points except in the limit of 
high frequencies. Here, the average mobility is temperature dependent and the 
interesting point is that the mobility decreases with temperature. This is a 
many-body effect and says that temperature is enhancing the coulombic 
disorder by allowing configurations which provide a wider distribution of activation
energies.

In order to understand this behavior more accurately, we monitored the motion of 
the charges in the column as a function of time. The main conclusion is that the 
charges follow the periodic pattern imposed by the external alternating field. 
When the field frequency is high, the charges do not have enough time to come 
closer to each other or to the electrodes, so they fluctuate rapidly around 
their initial positions following the field, and this gives rise to the given calculated 
mobility value shown in figures \ref{fig7} and \ref{fig8}. As the temperature increases, the thermal energy of 
the charges allows them to move more easily against the direction of the fields.
This increases the electron-electron disorder energy, and in this way
temperature now reduces the overall mobility! This picture explains the 
decrease of the mobility with temperature at a given value of (high) frequency, 
as shown in figure \ref{fig8}.

For a given high frequency range, this picture remains basically the same. 
Suppose that 3 charges are located in the column at some equilibrium positions. 
For a frequency 10$^{12}$ Hz, for example, they will make very short moves 
towards one or the other side. When the frequency is switched to 10$^{10}$ Hz, the 
charges will again follow the field, but despite the larger distances traveled 
in any one direction, the charge closest to the electrode will still not have 
enough time to seriously be blocked by the confining electrodes. Similarly, the other 
charges will not have enough time to feel the changes in each others positions, 
and they will continue to oscillate, essentially undisturbed, thus giving rise
to the same mobility as for higher 
frequencies. This explains the constant value of the $\mu(N)$ mobility versus
frequency at high frequencies.

When the frequency is small so that the displacement can reach the electrode boundary,
the charges are pushed towards one side of 
the column. They come very close to each other and thus the mobility decreases with $N$
(only the one furthest from the edge can move easily, but when the charge tries 
to escape, the external field pushes it back). As the frequency gets lower and lower, the 
effect becomes more pronounced because now the biasing field remains at the same 
magnitude and in the same direction for a longer time $1/\omega$. This is the reason why
the mobility falls rapidly and monotonically with lower frequencies.
In fact, when using blocking 
electrodes, the DC mobility is rigorously 0 as it should
be, with the (+) charges `piling' up near the (-) electrode.

\section{Conclusions}

We have continued the study of charge transport in interacting quasi-one-
dimensional gated systems with emphasis on the temperature dependence of the 
mobility. We have seen that the temperature changes the mobility in the presence 
of a DC field but that the effect is very much dependent on charge 
concentration. The temperature dependence is partially due to the assumed 
injection barrier and partially due to the fact that carriers can ``push their 
way in'' like a line of ``billiard balls''. Both factors work together to 
produce the final result shown in figure \ref{fig2}. Temperature can cause the 
net steady state charge to vary and this then also affects the charge-charge 
interaction efficiency  (figure \ref{fig2}).

The many body interactions do indeed cause a frequency dependence of the 
mobility. However, using the same parameters as for doped discotic HAT6\cite{18},
we find that the 
frequency dependence comes in at too high a frequency. It seems that the experimental data in 
references \onlinecite{1a,1b,1c,1d,18} can only be understood if we also include 
disorder. 

When we have blocking electrodes, the interactions influence the density 
and the temperature dependent ac mobility. The charges increase the confinement 
and the mobility decreases with concentration and with temperature in the way one would expect
for finite ordered segments.
Finally, we now need to extend our theoretical tools to handle 2-dimensional 
systems, so that we can model smectic liquid crystals, mixes of smectics, FET's, thin-film
transistors (TFT),
and also the transport through large molecules such as nanotubes and nanotubes filled with
fullerenes (pea pods)\cite{17}. So far the work has focused on quantum
transport. There are and will be, we believe, many new and exciting
experimental results coming up in this field in the stochastic and
diffusion  limit.

\begin{acknowledgments}
One of the authors, B. Movaghar, would  like to thank NEDO Int. Joint Research  
Grant for financial support. 
\end{acknowledgments}


\begin{thebibliography}{}

\bibitem{1a} N. Boden \emph{et al.}, Liq. Cryst. {\bf 15}, 851 (1993).

\bibitem{1b} N. Boden, R.J. Bushby, J. Clements, and B. Movaghar, J. Appl. Phys. 
{\bf 33}, 3207 (1995).

\bibitem{1c} K. Arikainen \emph{et al.}, J. Mat. Chem. {\bf 5}, 2161 (1995).

\bibitem{1d} N. Boden and B Movaghar, in \emph{Liquid Crystal Handbook}, edited 
by D. Demus, J.W. Goodby, G.W. Gray, H.W. Spiess, and V. Vill,
Vol. 2B, 781 (Wiley-VCH, 1998).

\bibitem{2a} D. Adam \emph{et al.}, Phys. Rev. Lett. {\bf 70}, 457 (1993).

\bibitem{2b} A. Bacher \emph{et al.}, Adv. Mater. {\bf 9}, 1031 (1997).

\bibitem{3} N. Boden \emph{et al.}, Phys. Rev. B {\bf 52}, 13274 (1995); {\bf 
58}, 3063(1998); {\bf 65}, 104204  (2002).

\bibitem{4} Z. Yao, H.W.Ch. Postma, L. Balents, and C. Dekker, Nature {\bf 402}, 
273  (1999).

\bibitem{5} D.L. Carroll \emph{et al.}, in Molecular nanostructures, edited by H. Kuzmany, J. Fink, M. 
Mehring, and S. Roth, 477 (World Scientific, Singapore, 1997);
Carbon {\bf 36}, 753 (1998).

\bibitem{6} M. Rother \emph{et al.}, Microelectronics Eng, {\bf 47}, 215 (1999).

\bibitem{7} M. Funahashi and J. Hanna, Appl. Phys. Lett. {\bf 76}, 2574  (2000).

\bibitem{8} M. Funahashi and J. Hanna, Mol. Cryst. Liq. Crystals {\bf 331}, 2369 
(1999).

\bibitem{9} J. Ribo \emph{et al.}, Synthetic metals {\bf 97}, 229 (1998).

\bibitem{10} L. Torsi \emph{et al.}, Sensors and actuators B {\bf 77}, 7 (2001).

\bibitem{12a} R.F. Service \emph{et al.}, Science, {\bf 283}, 1667 (1999).

\bibitem{12b} J. Chen, M.A. Reed, A.M. Rawlett, and J.M. Tour, Science {\bf 
286}, 1550 (1999).

\bibitem{12c} C. Dekker and M. Ratner, Physics World {\bf 14}, 29 (2001).

\bibitem{13} L.D.A. Siebbeles and B. Movaghar, J. Chem. Phys. {\bf 113}, 1609 
(2000).

\bibitem{14a} C. Zhou, D.M. Newns, J. Misewich, and P.C. Pattnaik, Appl. Phys. 
Lett. {\bf 70}, 598  (1997).

\bibitem{14b} T. Ouisse and T. Billon,  Phil. Mag. B, {\bf 71},  413 (1995).

\bibitem{15} H. B\"{o}ttger and V.V. Bryksin, \emph{Hopping Conduction in Solids}
(Akademie Verlag, Berlin, 1985).

\bibitem{16} S. Alexander, J. Bernasconi, and W.R. Schneider, Rev Mod Phys {\bf 53},
175 (1981); B. Movaghar, D.W. Murray, B. Pohlmann, and D. Wuertz,
J. Phys. C {\bf 17}, 1677  (1984).

\bibitem{18} N. Boden, R.J. Bushby, and J. Clements, J. Chem. Phys. {\bf 98}, 5920 (1993).

\bibitem{17} D.J. Hornbaker \emph{et al.}, Science, {\bf 295}, 828 (2002);
A. Rochefort, P. Avouris, F. Lesage, D.R. Salahub, Phys. Rev. B  {\bf 60}, 13824 (1999).

\end{thebibliography}
\end{document}